\documentstyle[prl,aps]{revtex}

\begin{document}
\twocolumn
\draft

{\bf Comment on ``Quantum Phase Slips and Transport in Ultrathin Superconducting
Wires''}

In a recent Letter \cite{ZGVZ}, Zaikin, Golubev, van Otterlo, and Zimanyi
(ZGVZ) criticized the phenomenological time-dependent Ginzburg-Laudau 
(TDGL) model which I used to study the quantum phase slip (QPS) for 
superconducting wires \cite{duan}. They claimed that they developed a "microscopic"
model, made {\it qualitative} improvement on my overestimate of the
tunnelling barrier due to electromagnetic field (EM), and claimed agreements
with the experiments by Giordano \cite{nick}. 

In this comment, I want to point out that, i), ZGVZ's 
result on EM barrier is expected in \cite{duan};
ii), their work is also 
phenomenological; 
iii), their renormalization scheme is fundamentally flawed; iv), 
they underestimated the barrier for ultrathin wires; 
v), their comparison with experiments
is incorrect. Details are given below.

i), In  \cite{duan} I emphasized results on relatively thick
wires and concluded that the observations on Giordano's
wires with thickness $\sqrt{\sigma}=$$410-505\AA$
must be due to weak links. Both kinetic inductance
and Mooij-Sch\"on mode have been
included \cite{duan,scot}. The ultrathin wire limit
of the Eq.(8) in  \cite{duan} gives an EM barrier,
\begin{eqnarray}
\frac{S_{EM}}{\hbar} \cong \frac{\pi}{2}
\sqrt{2\ln(q_{upper}/q_{lower})} \frac{\sqrt{\sigma}}
{\lambda_L} \frac{\hbar c}{e^2},
\label{ultrathin}
\end{eqnarray}
where the ratio of cutoffs $q_{upper}/q_{lower}\ge 10$.
{\it Eq.~(\ref{ultrathin}) is the main result in
Ref.\cite{ZGVZ}} except some underestimate (cf., below).

ii), QPS is a far-from-equilibrium process 
occuring on a time scale of
the order of inverse gap of the superconductors. 
ZGVZ's model contains nothing more than 
a saddle point plus Gaussian fluctuations 
and does not go beyond TDGL.
Despite minor differences from my work, Ref.\cite{ZGVZ} 
does not present any new physics. 
I also want to point out that their electric screening
for the superconducting charge is described by condensate fraction
$n_s$ which vanishes at $T_c$. This is incorrect since Debye screening
is nondiscriminating against superconducting or normal charge.

iii), ZGVZ's renormalization scheme 
is wrong since EM field {\it qualitatively} changes the Kosterlitz
renormalization flow \cite{scot,fisher,zhang}.
Therefore their conclusions on the ``metal-superconductor'' transition
and wire resistance were not based on a solid foundation.

iv), ZGVZ underestimated the EM barrier for
ultrathin wires. For Giordano's wire of
$\sigma=\pi(80\AA)^2$, if we {\it assume}
it is homogeneous with an effective London
penetration depth
$\lambda_L=1000\AA$, the EM barrier {\it alone}
reduces the tunneling probability to $e^{-60}$.
Taking into account the substrate dielectric constant
\cite{scot} $\epsilon=10$ for $Ge$, then the EM
suppression is $e^{-330}$. {\it In order to observe
QPS in homogeneous wires, the wire radius must
be smaller than $10\AA$}.

v), It needs to be emphasized that the results by 
Sharifi {\it et al.} \cite{dynes} are NOT
similar to that of \cite{nick}, but qualitatively different. 
The wires in \cite{dynes,neg}
are known to be homogeneous. In thinner wires
($[40\AA]^2$) than those in \cite{nick}, no quantum phase-slip was observed.
The wire resistance still obeys thermally activated behavior,
albeit being larger than predicted by classical models \cite{neg}.
As stated by ZGVZ, the theory is for a homogeneous wire. Hence they
should compare with the experiments in \cite{dynes} instead of
\cite{nick}. A random Josephson junctions model \cite{scot}
is suitable for granular wires \cite{nick,peng}.

TDGL was used to calculate the EM barrier for the 
quantum decay of supercurrent \cite{duan} and the EM contribution
to vortex mass \cite{mass}. 
There is a deep physical reason \cite{mass} for the second 
order time derivative term in TDGL: superfluids are compressible
due to number-phase conjugation. 
{\it Hence the qualitative physics in \cite{duan,mass}
is much more general than the formalism of the TDGL model}.

This work was supported by NSF (DMR 91-13631).

Ji-Min Duan

Department of Physics,
 
University of California-San Diego,
 
La Jolla, California 92093-0319



\begin{references}

\bibitem{ZGVZ} A. D. Zaikin, D. S. Golubev, A. van Otterlo,
and G. T. Zimanyi, 
Phys. Rev. Lett.
{\bf 78}, 1552 (1997).

\bibitem{duan} J.-M. Duan, Phys. Rev. Lett.
{\bf 74}, 5128 (1995).

\bibitem{nick} N. Giordano, Phys. Rev. Lett.
{\bf 61}, 2137 (1988); Physica B {\bf 203}, 460 (1994).

\bibitem{scot} S. R. Renn and J.-M. Duan, Phys. Rev. Lett.
{\bf 76}, 3400 (1996).

\bibitem{fisher} M. P. A. Fisher and G. Grinstein, Phys. Rev. Lett.{\bf 60}, 208 (1988).

\bibitem{zhang} S. C. Zhang, Phys. Rev. Lett.
{\bf 59}, 2111 (1987).

\bibitem{dynes} F. Sharifi, A. V. Herzog, and R. C. Dynes, Phys. Rev. Lett.
{\bf 71}, 428 (1993).

\bibitem{neg} P. Xiong, A. V. Herzog, 
and R. C. Dynes, Phys. Rev. Lett.
{\bf 78}, 927 (1997).

\bibitem{peng} A. V. Herzog, P. Xiong, F. Sharifi, 
and R. C. Dynes, Phys. Rev. Lett.
{\bf 76}, 668 (1996).

\bibitem{mass} J.-M. Duan, Phys. Rev. B {\bf 48}, 333 (1993);
{\it ibid}, {\bf 49}, 12381 (1994). J.-M. Duan, and A. J. 
Leggett, Phys. Rev. Lett.
{\bf 68}, 1216 (1992).


\end{references}
\end{document}